\documentclass[fleqn,10pt]{wlscirep}
\title{On the observability of Pauli crystals}

\author[1]{Debraj Rakshit}
\author[1]{Jan Mostowski}
\author[1]{Tomasz Sowi{\'n}ski}
\author[1]{Magdalena Za\l uska-Kotur}
\author[1]{Mariusz Gajda}
\affil[1]{Institute of Physics of the Polish Academy of Sciences, Al. Lotnik{\o}w 32/46, 02-668 Warszawa, Poland}

\begin{abstract}
The best known manifestation of the Fermi-Dirac statistics is the Pauli exclusion principle: no two identical fermions can occupy the same one-particle state.  This principle  enforces high order correlations in systems of many identical fermions and is responsible for a particular geometric arrangement of trapped particles even when all mutual interactions are absent \cite{Gajda16}. These geometric structures, called Pauli crystals, are predicted for a system of $N$ identical atoms trapped in a harmonic potential. They emerge as the most frequent configurations in a collection of single-shot pictures of the system.  Here we study  how fragile Pauli crystals are when realistic  experimental limitations are taken into account. The  influence of  the number of single-shots pictures available to analysis,  thermal fluctuations and finite efficiency of  detection are considered.  The role of these sources of noise  on the possibility of experimental observation  of Pauli crystals is shown and  conditions necessary for the detection of the geometrical arrangements of particles are identified.
\end{abstract}
\begin{document}

\flushbottom
\maketitle
%
%
\thispagestyle{empty}


\section*{Introduction}
\label{sec_introduction}
Ultracold atoms provide an ideal playground  for simulating not only various many-body quantum systems  \cite{Bloch08,Jaksch98,Greiner02,Jordens08,Lewenstein07,Lewenstein12}, but also for studies of few-body physics\cite{blume12}. Unprecedented progress in controlling and monitoring of ultracold atoms opens a whole world of new possibilities. For example  controlled few-body systems with sufficiently suppressed atom number fluctuations can be achieved in the Mott insulating phase of a lattice system via single-site addressing techniques \cite{Weitenberg11}. Alternatively, deterministic preparation of a few-body system is possible in a single microtrap   \cite{Serwane11,Wenz13}. Small systems of 1-10 fermionic atoms can be prepared in well-defined quantum states with high fidelity. 

Recently developed single-site fluorescence imaging technique is a major breakthrough in monitoring of many-body systems \cite{Bakr09,Bakr10,Sherson10,Cheuk15,Parsons15,Haller15,Edge15,Cheuk16}.  This great imaging techniques can simultaneously determine  positions of all individual atoms in optical lattices with single-site  resolution \cite{Bakr10,Sherson10}. Snapshots of the atomic ensemble within a single isolated trap could harbor many crucial information about the system. Very important in this respect are higher-order spatial correlations, which are elusive to one-particle detection, but can be revealed from the $N$-body probability density \cite{andrews97,Castin97,Hofferberth06,Schauss15}. 
One of the best examples is recent experiment \cite{Schauss15} in which the detection of spontaneous  self-organized ordering in Rydberg many-body systems  created in a Bose-Einstein condensate has been shown by comparing many single-shot pictures of the system.

Experimental progress challenges theory. Quantum many-body systems observed with single-particle resolution attract interest of physicists because they give access to information which is not accessible not only to the most common one-body measurements, but also to two-point correlations. Examples of processes whose understanding requires description going beyond two-point correlations are quite numerous. They include for instance the seminal paper   \cite{Javanainen96}, which shows the appearance of interference fringes in the course of $N$-body detection, no interference is seen in the one-body picture. Other prominent result  shows emergence of solitons from a  type II excited state of a one-dimensional system of bosons interacting via short-range potential described by the Lieb-Liniger model \cite{Syrwid15}.  The solitons appear as a result of high-order correlations in the system and can be uncovered in a single-shot detection of many atoms. The work in \cite{Sakmann16} has demonstrated how to simulate single-shot experiments of general ultracold bosonic systems based on numerical solutions of the many-body Schr{\"o}dinger equation. Simulations of time-dependent single-shot pictures monitoring  many-body dynamics revealing the appearance of fluctuating vortices in attractive BEC were suggested \cite{zoller16}. 
 
In this paper we investigate yet another example of high order correlations in a few-body system. The correlations we are interested in  are solely due to the quantum statistics and indistinguishability. Indistinguishable nature of identical particles leads very important consequences in quantum physics. It inherently imposes correlations between particles, even in absence of any mutual interactions. It  turns  out   that when many identical fermions at zero-temperature are trapped within an external two-dimensional harmonic trap, the particles orient themselves in a striking fashion unveiling specific geometric structures, called Pauli crystals. We have  shown recently how to extract  these geometric structures  from  multiple single-shot pictures of the many-body system \cite{Gajda16}. Suitable  analysis of a large number of data gives clear, unequivocal picture  of  Pauli  crystals. There is  a  question,  however, how much the observed images are deteriorated by various experimental imperfections and  if  the existence of Pauli crystal can  be  verified  experimentally. Very recently analogous Pauli structures by substituting Pauli principle by statistical interaction potential \cite{batle17}.

A single-shot measurement via fluorescence microscope can determine the spatial configuration of $N$ atoms. Detection of Pauli crystals from single-shot measurement's outcomes may, however, be challenging.  In a single-shot the positions of $N$-atoms can be obtained but  the outcome of such detection is unpredictable because of probabilistic nature of quantum mechanics. It is different in every realization. Therefore it is very difficult to notice any geometric arrangement of atoms.   But the most frequent ones might be quite similar  \cite{Gajda16}. In fact $N$-body probability density has a maximum for a certain arrangement of fermionic atoms -- Pauli crystals. One can expect that in a collection of $N$-particle pictures, there are many such results where atoms occupy positions "around" vertexes of a Pauli crystal. Uncertainties of the positions due to quantum shot-to-shot fluctuations causes the crystal vertexes to be spatially extended even at zero temperature. Moreover,  there are  several other practical constrains that might lead to additional smearing of crystal vertexes.  These are  thermal fluctuations, limited number of single-shot experiments, and  imperfections due to atom number fluctuation in the measured ensemble of single-shot pictures.  All these imperfections  may even cause the crystalline structures to be beyond the experimental reach. The present paper provides careful examination of a role of these fluctuations  to understand if Pauli crystals can be observed. We take into account practical constraints mentioned above.  It is shown below that  the  geometry of  the  Pauli  crystals  remains seen even if it is diffused by some amount of experimental noise. The noise however, has to be rather small.

\section*{The  model}
We consider non-interacting polarized fermions trapped in a two-dimensional isotropic harmonic potential with the frequency $\omega_x=\omega_y=\omega$. The single-particle quantum state, $\psi_{nm}(x,y)$, bound in the 2D harmonic trap is given by
\begin{equation}
\psi_{nm}(x,y)=\mathcal{N}_{mn}e^{-(x^2+y^2)/2}\mathcal{H}_n(x)\mathcal{H}_m(y),
\end{equation}
where $\mathcal{H}_n$ is the $n^{th}$ Hermite polynomial, $\mathcal{N}=(2^{n+m}n!m!\sqrt{\pi})^{1/2}$ is the normalization constant. Quantum  numbers $n$ and $m$ enumerate excitations in the $x$ and $y$ direction, respectively. We use natural harmonic oscillator units, $a_0=\sqrt{\hbar/M \omega}$, where $M$ is the mass of the particle. The single-particle energy corresponding to the quantum state $\psi_{mn}$ is $E_{mn}=(m+n+1) \hbar \omega$, and each of these energy levels are $(m+n+1)$-fold degenerate.

In the $N$-body ground state all identical fermionic atoms occupy the lowest available single-particle quantum states up to the Fermi energy. This is due to the Pauli exclusion principle, which prohibits them from occupying the same quantum state. The ground state is not uniquely defined if the number of states at the Fermi level exceeds the number of particles which have to be located there.   However, in  particular cases  when $N = 1, 3, 6, 10, 15, \ldots$ , the many-body ground state of the corresponding system is non-degenerate. Obviously, geometry of Pauli crystals is characteristic of a given state. If energy does not specify uniquely the state we have a variety of different states and geometric configurations corresponding to a given energy. Non-zero temperature resolves this issue. All states of the same energy contribute with equal weights to the thermal density matrix. Here, for a simplicity, we limit our discussion to isotropic trap and fully occupied Fermi surface.     

The ground state energy is the sum of single particles energies of the occupied states, $\mathcal{E}_{0}^N/{\hbar \omega}=\sum_i^N(m_i+n_i+1)$. Note that  any two given sets of  $(m_i,n_i)$ associated with a single-particle orbital cannot be identical since the many-body ground state wavefunction, $\Psi_{0 q}^N({\bf r}_1, {\bf r}_2, \ldots, {\bf r}_N)$,  is obtained by imposing anti-symmetrization via the Slater determinant of the occupied single-particle states.  The subscript $q$ in $\Psi_{0 q}^N$ accounts for the degeneracy of the N-body ground state.  The probability, $\mathcal{P}_q^N$, of finding  particles at $({\bf r}_1, {\bf r}_2, \ldots, {\bf r}_N)$  when the system is in the ground state is:
\begin{equation}
\label{eq_pn_0}
\mathcal{P}_q^N =  |\Psi_{0 q}^N({\bf r}_1, {\bf r}_2, \ldots, {\bf r}_N)|^2 .
\end{equation}

\begin{figure}[t]
\begin{center}
\includegraphics[angle=0,width=7.0cm,height=7.0cm]{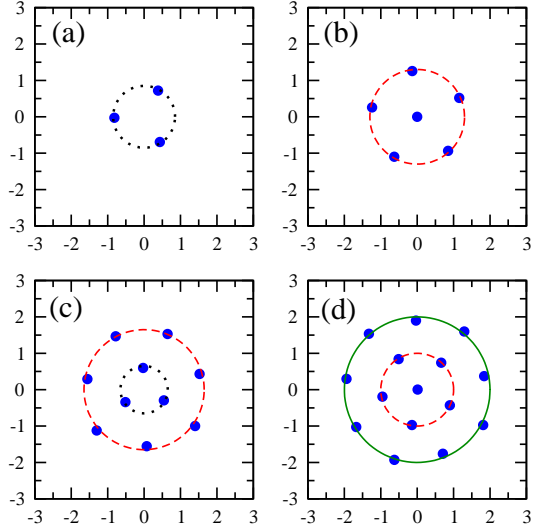}
\end{center}
\caption{\emph{Crystalline structures formed by identical fermions in two-dimensional harmonic trap.} The blue dots show the most probable configurations obtained by maximizing the $N$-particle probability for (a) 3 atoms,  (d) 6 atoms, (e) 10 atoms, and (f) 15 atoms. The black dotted, the red dashed, and the green solid circles represent the first shell (innermost), second shell, and third shell, respectively. First shell reduces to a point, when it is  composed  of  a single atom present  in the  center.}
\label{fig1}
\end{figure}

On a theory ground  the maximum of $\mathcal{P}_q^N$, Eq.~(\ref{eq_pn_0}), can be easily obtained. For example it can be found by employing the Monte-Carlo algorithm \cite{Metropolis53}, where starting from a randomly chosen initial configuration, another configuration is proposed by shifting the atom positions. The move is accepted if the  configuration is more probable than the previous one.

It turns out that, as it has been shown in \cite{Gajda16}, the most probable arrangement of atoms manifest unique geometrical structures at zero temperature. These configurations for  $N$ = 3, 6, 10, and 15 are shown in Fig.~\ref{fig1}. Three  atoms orient themselves at the vertexes of an equilateral triangle. A two-shell structure is noticed for $N$ = 6 with a single atom positioned at the trap center, representing the first (innermost) shell, and the second shell formed by five atoms arranged in a  pentagon. $N=10$ configuration exhibits a two-shell structure with a triangle and a heptagon, respectively, forming the inner and outer shells. A third shell emerges for $N=15$ atoms, where one atom occupies the trap center, and equally spaced five and nine atoms form the second, and the third (outermost) shell. It can be seen that the relative orientations of the shells are rigid with respect to each other. However  each structure as a whole can  be  rotated without affecting the value of $\mathcal{P}_q^N$.

\section{Image  processing}

We assume here that positions of atoms occupying the harmonic trap can be determined with high resolution, larger then their separation which is of the order of oscillator unit of length. This assumption is quite optimistic at this moment. At present the resolution of atomic fluorescence microscopes is higher than the lattice spacing and one can distinguish atoms at different sites. It is not possible however, to resolve positions of atoms residing in the same site. To overcome this problem atoms ought to be released prior to detection of their positions. Only  after some expansion the atoms can be distinguished. Ballistic expansion however, does not change the geometry of the initial arrangement. For the harmonic confinement expansion does not disturb geometry of the system, it results in  scaling of all distances  and we ignore it in the following analysis.

As an outcome of an instantaneous picture of all atoms in their ground state we get a set of $N$ position vectors. But this outcome is totally unpredictable due to the probabilistic nature of quantum mechanics. Consequently, configurations emerging out of different realizations are different and each of them is unpredictable. Therefore a single picture of $N$-body system  cannot reveal the geometric configurations of particles predicted by using the probability distribution. The natural strategy could be to repeat the experiment with the same initial state and combine all outcomes of the measurements to form a histogram of atomic positions in 2D-plane. Surprisingly enough, the histogram of repeated measurements does not help too. It gives a one particle density, not  the $N$-particle correlation function. 

Indeed, the histogram of $L$ simultaneously repeated measurements of positions ${\bf x}_i^s$ of $N$ atoms is defined as:
\begin{equation}
\label{H(x)}
H({\bf X})= \frac{1}{L}\sum_s \sum_{i=1}^{N}\delta({\bf x}^s_i-{\bf X}),
\end{equation}
where index $s$ refers to different measurements. The function  $\delta({\bf x}-{\bf X})$ is equal to one if  the position ${\bf x}$ of a particle coincides with the position of  a detector located at ${\bf X}$ with an accuracy $\Delta$, i.e.  if a  particle is found in a volume $V=\Delta^2$ around ${\bf X}$.  In an opposite case, the function $\delta({\bf x}-{\bf X})$ vanishes. Obviously, $H({\bf X})$ is proportional to  a quantum-mechanical  one-particle probability distribution. This can be seen by changing the order of summation. The histogram has no information about  correlations of particles' positions because summation over configurations can be done independently for every particle.  This  erases information about relative positions of particles and a geometry the system.  

If we have no information about the system at hand as well as about its  symmetries, then we have to refer to conditional probabilities.  Prior to processing the outcomes of measurement it is useful to shift every configuration and locate its geometric center  at the center of coordinates system.  This way the quantum uncertainty of the center of mass position is eliminated. The algorithm of finding  the most probable configuration could be  divided into three steps then. (i) Having collection of single shot pictures one has to find histogram of configurations and  determine its maximum. (ii)  Next, all  single-shot pictures with no particle  close (within some predefined distance) to the maximum  should be removed from further analysis  (iii) The procedure have to be repeated with the reduced set of pictures  starting from (i) until  all $N$  maxima  are found.  In case when there are several local maxima one should choose a global maximum if such exists, or any of the equivalent maxima, in the opposite case. 

\begin{figure}[t]
\hspace*{-1.0cm}
\begin{center}
\includegraphics[angle=0,width=10.0cm,height=3.5cm]{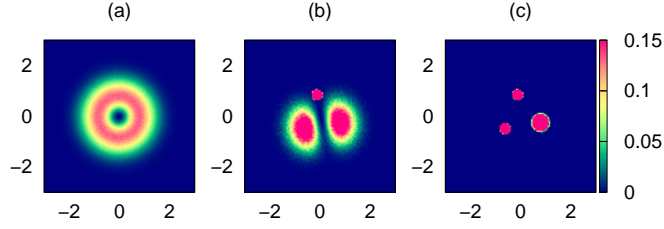}
\end{center}
\vspace*{-1.0cm}
\caption{\emph{Conditional probability densities.} (a) one-particle density after removing the center of mass uncertainty, (b) two-particle conditional probability density, (c)  three-particle conditional probability density.}
\label{xxx}
\end{figure}

 To illustrate this procedure based on conditional probabilities, in Fig.~\ref{xxx} we show results of consecutive steps described above for $N=3$ atoms. In the first panel (a) the one particle probability density, after eliminating the  center of mass uncertainty, is shown. Because of this elimination the density has a minimum at the center and reaches a maximal value on a ring. Then we select one point on the ring and keep only the pictures in which there is a particle at a distance smaller then  $R$ from the selected point. Obtained this way conditional probability density is shown in panel (b). Two equal maxima are visible. In the third panel (c) we show the conditional probability obtained by removing from panel (b) all these configurations which do not contribute to the neighborhood of the two previously found maxima.  The Pauli crystal -- configuration maximizing the three-body probability density, can be seen. 

The above procedure is very wasteful if the system has some symmetries. Many  measurements might be removed from a collection of pictures only because they differ by a symmetry transformation. In Fig.~\ref{xxx}(a) some $L=10^7$ snapshots are taken, while the procedure leading to   Fig.~\ref{xxx}(c) left only about $7\times10^4$ snapshots.
The system we study has some symmetry and we will use another procedure of image processing\cite{Gajda16}.
Let  us note  that in every measurement the rotational symmetry of the one-particle density is broken differently. Each measured configuration represents a whole class of configurations which differ by a symmetry transformation only. We want to take  advantage of these symmetries and  compare all single-shot pictures, to the one maximizing the $N$-body probability. Here we assume that knowing the system at hand we know the geometrical arrangement at zero temperature, i.e., the geometry of the Pauli crystal, the vertexes of which are assumed to be positioned at $\{{\bf r}_0\}_N= \{{\bf r}_{01},{\bf r}_{02}, \ldots, {\bf r}_{0N}\}$. The latter can be obtained on the theoretical ground. All single-shot outcomes $\{{\bf x}\}_N=\{{\bf x}_1,{\bf x}_2, \ldots, {\bf x}_N\}$ corresponding to a particular measured realization are compared then to the pattern after some previous preprocessing - a symmetry transformation.

As we will not pay attention to a particular enumeration of particles, the only relevant symmetry operation in our case is rotation around the trap center.  To eliminate the quantum uncertainty of the center of mass and to concentrate on the relative positions of particles we shift the center of mass of every configuration to the origin of the coordinate system, ${\bf x}'_i={\bf x}_i-{\bf x}_{cm}$  where ${\bf x}_{cm}=(1/N)\sum_{i=1}^N {\bf x}_i$. For simplicity we skip the `{\it prime}' symbol in the following.

Quantitative comparison of a particular snapshot to the pattern has to be based on a definition of the distance between the snapshot and the pattern. We will simply use the angular distance (squared) between particles of the pattern and the snapshot. It works well in our case. To every snap-shot particle at ${\bf x}_i$ we first assign its unique  partner in the pattern,  ${\bf x}_i \mapsto {\bf r}_{0\sigma (i)}$ by ordering both the pattern and the snapshot according to the increasing azimuthal angle of particles' coordinates within every shell separately. By ${\sigma}$ we denote the permutation of indices enumerating particles. At this step we use polar coordinates, ${\bf x}_i = (d_i,\phi_i)$, and ${\bf r}_{0i} = (\rho_i,\beta_i)$. After such ordering, assigning of the snap-shot particles  to the vertexes of the pattern is straightforward: the first particle from the snapshot is assign to the first in the pattern, the second to the second, etc.  The distance between the given configuration and the pattern is then:
\begin{equation}
\Theta^2( \{ {\bf x} \}_N)=\sum_{i=1}^{N}\left(\phi_i - \beta_{\sigma (i)}\right)^2.
\end{equation}

Single snapshot breaks the rotational symmetry of the system, in a different way in  every realization. Therefore we rotate every single shot configuration by an angle $\alpha$ in the $x-y$ plane,  ${\bf x}_i(\alpha)=\mathcal{R}_{\alpha}({\bf x}_i)$ to minimize the distance between the snapshot and the pattern. Thus, the optimal angle of rotation, $\alpha_{\text{opt}}$ is given by:
\begin{equation}
\alpha_{\text{opt}}=\min_{\alpha}\{ \Theta^2(\{{\bf x}(\alpha)\}_N) : \alpha \in [0,\alpha_{\max}] \},
\end{equation}
where $\mathcal{R}_{\alpha}$ represents the rotational operator. The maximal rotation angle, $\alpha_{\max}$, is fixed to be $2 \pi/k$ for the system with $k$-fold axis of symmetry. Similar preprocessing was used in \cite{Schauss15} to show a geometric arrangement of Rydberg atoms in a condensate.

The preprocessing described above attempts to fit every single-shot picture of the system to the Pauli crystal. The transformation used does not change the geometry of the system. The only result of the preprocessing procedure is to transform the coordinates of particles, from the original  $\{ {\bf x}^s \}_N$ to the final $\{ {\bf x}^s (\alpha^s_{opt}) \}_N$. The histogram of these optimally adjusted configurations, $C({\bf X})$:
\begin{equation}
\label{C(x)}
C({\bf X})= \frac{1}{L}\sum_{s=1}^L \sum_{i=1}^{N}\delta({\bf x}_i^s(\alpha^s_{opt})-{\bf X}),
\end{equation}
preserves all information about correlations between individual particles. Index $s$ refers to different measurements.

The  procedure described  above  takes  advantage  of  the system symmetries. It is very effective and allows to use  all   available data in the final  analysis, Eq. (\ref{C(x)}). 

\section{Role of experimental imperfections}

No experiment is perfect. We have identified several sources of noise that can influence the visibility of Pauli crystals and even destroy completely the possibility of detection. First of all any experiment can give only a finite number of single shot outcomes. Evidently the larger the number of repetitions of the experiment, the histogram of configurations detected becomes closer to the theoretical predictions based on quantum mechanics. The noise related to a finite number of measurements is related to statistical nature of the theory.

Finite temperature is another source of noise. If the system is in a thermal state with temperature above zero then some of the atoms occupy excited state of the trap with probability determined by the temperature. All states contributing to the thermal density matrix of the system have different geometry. Non-zero temperature leads to configuration-``mixing" of different geometries with thermal weights. Thermal fluctuations cause flattening of the probability distribution of different configurations, and consequently, non-zero temperature leads to additional spreading of the  particle' positions. 

Another source of noise comes form imperfect detection. Previous discussion assumed that the detection is ideal  --  all particles are  detected in every single shot measurement with $100\%$ efficiency. Evidently this is an idealization, in a real experiment we expect to find different numbers of particles in a collection of single shot pictures. There are two main sources of these shot to shot fluctuations. One stems from imperfect detection efficiency, the second one is a result of fluctuations of particle number in the system. The number of particles varies form shot to shot due to the destructive character of measurement and imperfect preparation of the initial state. Post selection has to be applied in order to keep the desired number of atoms $N$ in each analyzed picture. But even then we cannot be sure that initial system was composed of $N$ atoms. Imperfect detection can add admixture of configurations that are different from the ground state configuration. In this respect this is similar to the effect of non-zero temperature.

\subsection{Number of single-shot experiments}
\label{ftpc}
We start with analysis of the influence of finite number of single shot outcome on the visibility of Pauli crystals. We assume that the histogram $C({\bf X})$ (Eq.~(\ref{C(x)})) - the configuration probability density represents a typical output of the experiment. The configuration probability density is a result of $L$ independent measurements of positions of $N$ atoms  performed on a system trapped in a two-dimensional harmonic trap. Its contrast strongly depends on the number of single-shot pictures subject to analysis. 

\begin{figure}[t]
\hspace*{-0.5cm}
\begin{center}
\includegraphics[angle=0,width=10.0cm,height=10cm]{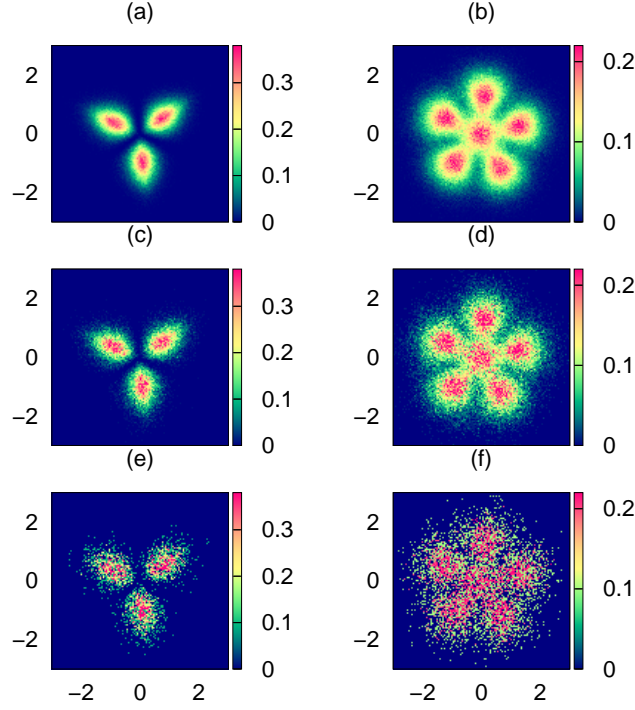}
\end{center}
\caption{\emph{Role of number of single-shot outcomes.} Visibility of Pauli crystals at $T=0$ for various number of single snap-shots analyzed. The left and the right panel shows configuration probability, $C({\bf X})$, for $N=3$ and $N=6$ atoms  for various number of snap-shots (a--b) $L=10^5$, (c-d) $L=10^4$, and (e-f) $L=10^3$.}
\label{fig2}
\end{figure}

Let us first look at the zero temperature case. In Fig.~\ref{fig2} we show the histogram of configurations $C({\bf X})$ for $N=3$ and $N=6$ atoms for various number of snapshot pictures $L$.  For large $L$ the structures visible there can be immediately compared with the most probable configurations obtained by maximizing the $N$-particle probability (see Figs.~(1.a) and (1.b)). Although quantum uncertainty causes considerable smearing of the  crystal vertexes, the single-shell Pauli crystal structure for three atoms and the double-shell structure for six atoms are clearly manifested. Smearing of the vertices signifies shot-to shot deformations of the observed configurations but  the underlying geometry is preserved. One can see that the contrast in the figure substantially deteriorates with decreasing number of configuration analyzed.  For  $L=1000$ which from experimental point of view seems to be  the most realistic case, some signatures of  geometric configurations are still noticeable, although the resulting histogram of  $C({\bf X})$ shows huge point-to-point fluctuations.  It varies a little from one realization to the other. 

\subsection{Temperature}
At finite temperature individual particles can access higher energy levels due to thermal fluctuations.  Let us denote the $N$-body energies by $\mathcal{E}_{\alpha}^N$, where $\alpha$ enumerates  consecutive eigen-energies. The many-body state $\Psi_{\alpha q}^N({\bf r}_1, {\bf r}_2, \ldots, {\bf r}_N)$  is obtained by imposing anti-symmetrization via Slater determinant of the occupied single particle orbitals.  The subscript $q$ in $\Psi_{\alpha q}$ accounts for the degeneracy of the $\alpha^{th}$ N-body energy state. The probability, $\mathcal{Z}_(N,T)$, of finding  particles at ${\bf r}_1, {\bf r}_2, \ldots, {\bf r}_N$ at a finite temperature $T$ is given by
\begin{equation}
\label{eq_pn}
\mathcal{Z}(N ,T) = \frac{ \sum_{\alpha} e^{-\mathcal{E}_{\alpha}^N/\kappa_{B} T} \sum_{q} 
|\Psi_{\alpha q}({\bf r}_1, {\bf r}_2, \cdots, {\bf r}_N)|^2 }{\sum_{\alpha} d_{\alpha} e^{-\mathcal{E}_{\alpha}^N/\kappa_{B} T}},
\end{equation}
where $d_{\alpha}$ is the number of degenerated states corresponding to the ${\alpha}^{th}$ energy level, and $\kappa_B$ is the Boltzmann constant. At low temperatures, the contribution to $\mathcal{P}_N$ from highly excited states is negligible. Consequently, for practical purpose of numerical computation, we adjust the cut-off in the weighted sum running over the index $\alpha$ in Eq.~(\ref{eq_pn}). In this work, we consider all  many-body eigenstates below the cut-off energy, $\epsilon_c$, which is set to $\epsilon_c=\mathcal{E}_{0}^N+6 \hbar \omega$, where $\mathcal{E}_{0}^N$ is the many-body ground state energy.
\begin{figure}[t]
\begin{center}
\includegraphics[angle=0,width=9.5cm,height=12cm]{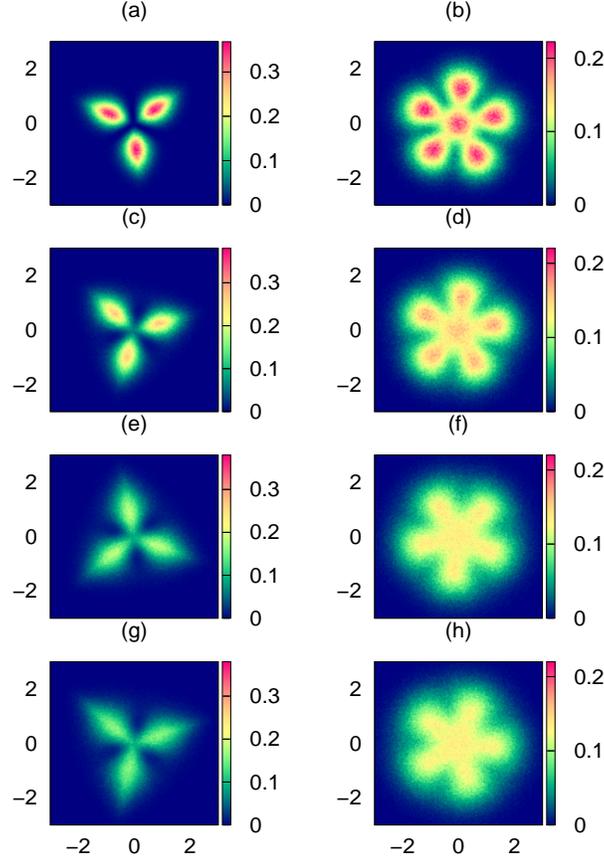}
\end{center}
\caption{\emph{Role of temperature. Configuration probability densities, $C({\bf X})$, at different temperatures.} The left and right panel, respectively, shows configuration probability densities for $N=3$ and $N=6$ atoms. Temperature grows from top to bottom: (a--b)  $\kappa_B T/{\hbar \omega}=0$ , (c--d) $\kappa_B T/{\hbar \omega}=0.5$, (e--f) $\kappa_B T/{\hbar \omega}=1.0$, and (g--h) $\kappa_B T/{\hbar \omega}=1.5$. Position is measured in natural units of the harmonic oscillator. The figure has been obtained with $L=10^6$ configurations.}
\label{fig3}
\end{figure}
As the temperature increases, the crystalline structures are getting `deformed' not only because of quantum uncertainty of particles' positions but also because of thermal fluctuations resulting from `admixtures' of contributions of excited states to the many-body density matrix, Fig.~\ref{fig3}. High order spatial correlations in excited states are different from those in the ground state, so the geometry of the configuration maximizing the probability is also modified. All these effects result in an increase of the spatial extensions of the Pauli crystal vertexes or in a change of their geometry, finally in total melting of the structures.   Fig.~\ref{fig3} shows $C({\bf X})$ for $N$ = 3, and $N$ = 6 at $\kappa_B T/\hbar\omega = 0, 0.5, 1.0$, and $1.5$. Although the single-shell Pauli crystal structure for three atoms remains visible at finite temperatures considered here, the distinction between two shells is blurred for $N=6$ at finite temperatures ($\kappa_B T/\hbar\omega>0.5$), as thermal fluctuations cause significant overlapping between atom positions from the inner and outer shells. The outer shell structure, however, can still be resolved for $N=6$ as the uncertainty in atomic positions remain smaller than the inter-atomic separation.
\begin{figure}[t]
\begin{center}
\includegraphics[angle=0,width=60mm,height=80mm]{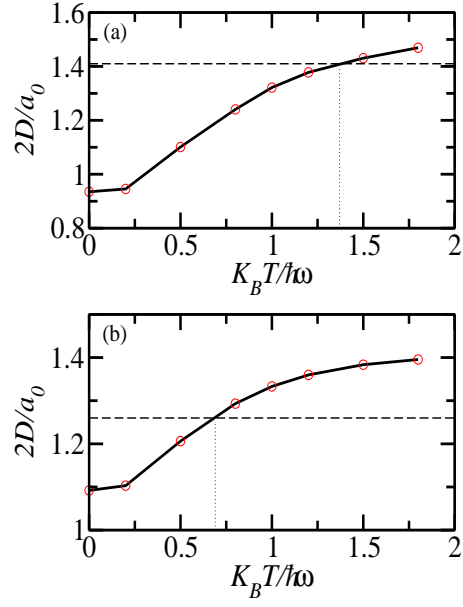}
\end{center}
\caption{(a) Diameter of the vertex position uncertainty  $2D$,  as a function of temperature, $T$ measured in units of the natural length of the harmonic trap, $a_0$. Horizontal line corresponds to a distance between vertexes of Pauli crystal for (a) $N=3$ atoms, (b) $N=6$ atoms. $L=10^6$.}
\label{fig4}
\end{figure}

As a result of thermal fluctuations the positions of atoms form a kind of finite size spots around vertexes maximizing the probability, see Figs.~(\ref{fig2}, \ref{fig3}). Characteristic size of these spots can be  estimated by the average distance of the atoms  located at ${\bf x}_i (\alpha_{opt})$ to the nearest vertex of the pattern situated at ${\bf r}_{0,i}$, i.e.  $\mathcal{D}\left({\bf x}_i(\alpha_{opt}),{\bf r}_{0,i}\right)$. We average the distance over all single shot configurations and all particles of the system:
\begin{equation}
D=\frac{1}{L} \sum_s \left( \frac{1}{N} \sum_i^N \mathcal{D}\left({\bf x}_i(\alpha_{opt}),{\bf r}_{0,i}\right) \right),
\end{equation}
The mean-distance monotonically increases with temperature for $N$ = 3 and 6 (see Fig.~\ref{fig4}(a)). In this figure we show the `diameter' of the vertex spot, $2D$, as a function of temperature.  By a horizontal line we show characteristic separation between vertexes of the Pauli crystal. Melting temperature can be estimated as the temperature at which the separation between vertexes is equal to the diameter of the spot area around the vertex.

In a case of $N=3$  this temperature is about $T_c(N=3) \approx 1.38 \hbar \omega$. Note however, that the area occupied by atoms around the vertex of the Pauli crystal is strongly elongated in the radial direction, so the average diameter $2D$ is a kind of average linear extension of the vertex.  The geometric structures for $N=3$ are still well resolved at temperatures larger than $T_c(N=3)$, Fig.~\ref{fig3}. In the case of $N=6$ particles, the spots have a more regular shape. The distances between atoms in the outer shell are larger then their distance to the center. Therefore, with increasing temperature the outer shell structure is still visible while the inner shell (in this case reduced to one atom only)   melts and the central spot overlaps with outer shell spots. The horizontal line in Fig.~\ref{fig4}), indicating the radius of the outer shell, crosses the diameter of the spot at the melting temperature $T_c(N=6) \approx 0.7 \hbar \omega$. At this temperature only the outer shell structure of the Pauli crystal can be visible, while the inner shell is smeared out. These simple estimations show that for larger number of atoms the Pauli crystal structures are not well resolved and to a large extend are hidden by thermal fluctuations. 

\subsection{Atom  number  fluctuation}
Previous discussion assumed perfect detection --  all particles are  detected in every single shot measurement with $100\%$ efficiency. Evidently this is  an idealization, in what follows we will consider the effect of an imperfect detection on images obtained. In fact in the real system we might expect to find a different number of particles in a collection of single shot pictures. These shot to shot fluctuations result both from imperfect detection efficiency and from fluctuations of the initial state.  Each measurement destroys the system, repetition of  experiment requires preparation of the initial state again. This preparation is not ideal, the number of atoms can differ in various realizations.  Direct inspection of pictures and post selection allows to keep for further analysis only these pictures in which the desired number of atoms $N$ is visible. But it does not mean that initial system was composed of $N$ atoms.  If the probability of {\it not-detecting} a particle is $\eta$, (detection efficiency is $1-\eta$) the observed configuration of $N$ atoms is randomly selected from  the following probability density distribution:
\begin{equation}
\label{QN}
\mathcal{Q} (N,T)=\frac{1-\eta}{1-\eta^{N_{m}+1}}\sum_{N_1=0}^{N_{m}} \eta^{N_1}\mathcal{Z}_T(N|N+N_1),
\end{equation}
where $\mathcal{Z}_T(N|N+N_1)$ is a reduced $N$-particle probability density of $N+N_1$-body system at temperature $T$ :
\begin{equation}
\mathcal{Z}_T(N|N+N_1)=\int \mathcal{Z}(N+N_1,T)
{\rm d}{\bf r}_{N+1} \ldots {\rm d}{\bf r}_{N+N_1},
\end{equation}
where $N_m$ is a maximal number of atoms in the system. A single shot picture of $N$-atoms might not correspond a particular realization of $N$-body system, but can represent a $N$-body subsystem of $(N+N_m)$ atoms. In a picture $N_m$ atoms is simply missing. We will not consider general case here but assume that the detection efficiency is  close to one, $1-\eta=0.9$. Therefore we can limit the summation in Eq.~(\ref{QN}) to $N_m=1$, i.e. we assume that at most one atom is missing in the picture, and  with the probability $1/(1+\eta) \approx 0.91$ the studied system was composed of $N$ atoms, and with probability $\eta/(1+\eta) \approx 0.09$ our system was composed of $N+1$ atoms:
\begin{equation}
\mathcal{Q} (N,T)=\frac{1}{1+\eta} \left[ (\mathcal{Z}_T(N|N)+\eta \mathcal{Z}_T(N|N+1)\right],
\end{equation}

\begin{figure}[t]
\begin{center}
\includegraphics[angle=0,width=9.5cm,height=4cm]{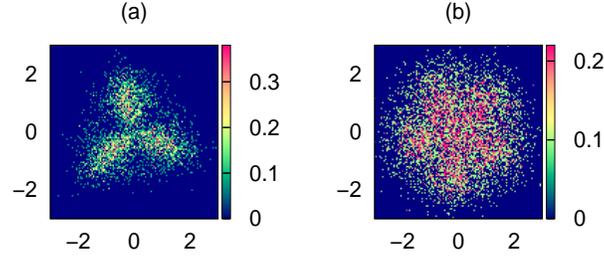}
\end{center}
\caption{\emph{Detection imperfection.} Configuration probability density, $C({\bf X})$, for imperfect efficiency $\eta=0.9$, non-zero temperature $\kappa_B T/\hbar \omega = 1.0$, and small number of snapshots $L=10^3$ for (a) $N=3$ and (b) $N=6$. Position is measured in natural units of the harmonic oscillator.}
\label{fig7}
\end{figure}

We  studied numerically the role of finite detection efficiency using example of three and six particles. In all cases the algorithm described above was used to fit detected particles positions to the pattern of 3 or 6 particles respectively. In Fig.~\ref{fig7} we present two cases, with zero temperature and finite temperature $T$ such that $\kappa_B T/\hbar\omega=1$. The number of single shots was rather small, $10^3$. Figure \ref{fig7} shows that the Pauli crystal structure is robust to  fluctuations caused not only by the temperature but to the particle number noise as well. We assumed high efficiency of detection, having in mind fluorescence microscopes used for this purposes. The fluorescence signal is rather strong as it results from a resonant process and scattering of about tens-hundred of photons. For the detection efficiency considered here the admixture of configurations corresponding to different number of particles is small and does not lead to essential modifications of the observed structures. It contributes to  noise, however, in the range of parameters of  interest, the most important contribution to the noisy background comes from thermal fluctuations.

\section*{Discussion}

We studied here possibilities of detecting the geometric structure formed by a small number of identical fermions. Because of their high symmetry the structures are called Pauli crystals. We  concentrated on the robustness of these structures in order to verify if they can be detected in real experiments. In order to do this we  included such factors as non-zero temperature and finite detection probability. Obviously both factors, as well as all other possible noise reduce the visibility of the structures. Our calculations impose quite stringent but realistic limitations on the noise: temperature smaller than $0.5\hbar\omega$ and efficiency of detection not worse then $0.9$. We show that under such conditions a geometry of Pauli crystals remains seen even when being diffused by noise.  Visibility of Pauli crystals strongly depends on the number of single-shot pictures available for pre-processing. At least $10^3$ copies is needed, however the larger the number the visibility gets better. 

It should be noted that direct observation of particle positions in case of particles in an optical trap is not possible. This does not depend on the particle statistics or whether they form a Pauli crystal or other geometrical structure. The reason is that distances between the particles are too small to be resolved by a direct observation.  The way to measure particle positions is to allow for free expansion of the particles following removal of the trapping potential. Individual particles are detected after sufficiently long expansion. In this way the measurement of particle positions reduces in fact to measurement of particle momenta in the original configuration. The transformation between the original position distribution and measured momentum distribution does not introduce noise, and therefore the Pauli crystal should be visible both in the position and momentum distribution.

Our calculations  show the level of experimental difficulties on the way to observe Pauli crystals. We hope that they  will encourage experimental groups to face the challenge.

\section*{Methods}

For generating ensemble of random configurations according to their $N$-body density profile, we use the Metropolis algorithm \cite{Metropolis53}. The random configurations are picked from a random Markov walk in the configuration space. Let $p$ be the transition probability defined as the ratio of probabilities of a trial configuration, $\{{\bf Y}\}$, and a given configuration, $\{{\bf X}\}$: $p=\mathcal{P}(\{{\bf Y}\})/\mathcal{P}(\{{\bf Y}\})$, where $\mathcal{P}$ is $N$ body probability distribution, and configurations $\{{\bf X}\}$ and $\{{\bf Y}\}$ describe positions of $N$ particles. The trial configuration is accepted if it is more probable, $p>1$. If $p<1$, the member of the ensemble is chosen probabilistically, it is the new or the old one. Decision is made,  depending on the value of random number $r \in [0:1]$ selected from the uniform probability distribution.  The trial configuration is accepted to the ensemble if $r<p$,   the "old" configuration is included again  if $r>p$. The members of this Marokv chain are subject to the probability distribution  $\mathcal{P}$ in the limit of infinite chain. Finite chains converge towards the probability $\mathcal{P}$, the rate of convergence is the best if on average every second trial configuration is accepted to the chain.


\section*{Acknowledgements}

The authors thank Immanuel Bloch for discussion. MG and DR acknowledge support from the EU Horizon 2020-FET QUIC 641122. TS acknowledges financial support from the (Polish) National Science Center Grant No. 2016/22/E/ST2/00555.







\end{document}